\documentclass[11pt,a4paper]{article}
\usepackage[hyperref]{naaclhlt2019}
\usepackage{times}
\usepackage{latexsym}
\usepackage[]{algorithm2e}

\RequirePackage{graphicx}

\usepackage{url}

\aclfinalcopy % Uncomment this line for the final submission
\title{Evaluation and Improvement of Chatbot Text Classification Data Quality Using Plausible Negative Examples}

\author{Kit Kuksenok \\
  jobpal Ltd. \\
  Berlin, Germany \\
  {\tt kit@jobpal.ai} \\\And
  Andriy Martyniv \\
  jobpal Ltd. \\
  Berlin, Germany \\
  {\tt andriy@jobpal.ai} \\}

\date{}

\begin{document}
\maketitle
\begin{abstract}

We describe and validate  a metric for estimating multi-class classifier performance based on cross-validation and adapted for improvement of small, unbalanced natural-language datasets used in chatbot design. Our experiences draw upon building recruitment chatbots that mediate communication between job-seekers and recruiters by exposing the ML/NLP dataset to the recruiting team. Evaluation approaches must be understandable to various stakeholders, and useful for improving chatbot performance. The metric, \texttt{nex-cv}, uses negative examples in the evaluation of text classification, and fulfils three requirements. First, it is actionable: it can be used by non-developer staff. Second, it is not overly optimistic compared to human ratings, making it a fast method for comparing classifiers. Third, it allows model-agnostic comparison, making it useful for comparing systems despite implementation differences. We validate the metric based on seven recruitment-domain datasets in English and German over the course of one year.

\end{abstract}
\section{Introduction}

\begin{figure}
    \includegraphics[width=220px]{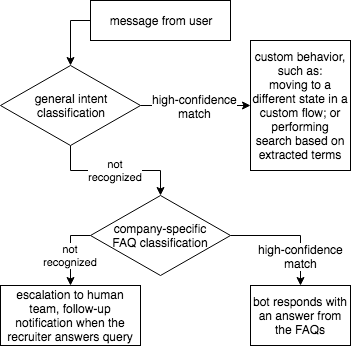}
    \caption{Each incoming message from an end-user is subject to (1) a  general intent classifier specific to  a language; and, if none of the roughly  20 intents are the recognized, (2) a company-specific FAQ classifier. Custom flow affects the specifics of this behavior.}
    \label{fig:flowcontrol}
\end{figure}

Smart conversational agents are increasingly used across business domains ~\cite{jain2018evaluating}. We focus on recruitment chatbots  that connect recruiters and job-seekers. The recruiter teams we  work with are motivated by reasons of scale and accessibility to build and maintain chatbots  that provide answers to frequently  asked questions (FAQs) based on ML/NLP datasets. Our enterprise clients may have up to $100K$ employees, and commensurate hiring rate. We have found that almost $50\%$ of end-user (job-seeker) traffic occurs outside of working hours  \cite{jobpaltiming2018}, which is consistent  with  the anecdotal reports of  our  clients that using the chatbot  helped reduce email and ticket inquiries of common FAQs. The usefulness of these question-answering conversational UIs depends on building and maintaining the ML/NLP components used in the overall flow (see Fig.~\ref{fig:flowcontrol}).

%The quaThe construction of small, natural-language datasets to anticipate domain-specific uses of a chatbot is a typical step in existing chatbot platforms \cite{canonico2018comparison}. Model-agnostic and unbiased evaluation and classifier or data error diagnosis is necessary for chatbot design, development, and maintenance.

In practice, the use of NLP does not improve the experience of many chatbots~\cite{pereira2018quality}, which is unsurprising. Although \textit{transparency} (being  ``honest and transparent when explaining why something doesn't work'') is a core design recommendation \cite{dialogflowerrors}, the most commonly available higher-level platforms~\cite{canonico2018comparison} do not provide robust ways to understand error and communicate  its implications. \textit{Interpretability} is a  challenge beyond chatbots, and is a prerequisite for trust in both individual predictions and the overall model~\cite{ribeiro2016should}. The development of the \texttt{nex-cv} metric was driven by a need for a quantification useful to  developers, as well as both vendor and client non-developer staff.%Usefulness combines interpretability, and capacity to enable these stakeholders to work together to understand  and improve  overall chatbot performance.

The  \texttt{nex-cv} metric uses plausible \underline{\textbf{n}}egative \underline{\textbf{ex}}amples to perform actionable, model-agnostic evaluation of text classification as a component in a chatbot system. It was developed, validated, and used at  \textit{jobpal}, a recruiting chatbot company, in projects where a client company's recruiting team trains and maintains a semi-automated conversational agent's question-answering dataset. Use of ML and NLP is subject to conversation flow design considerations, and internal and external  transparency needs \cite{kuksenok2019transparency}. The chatbots do not generate answers, but provide all responses from a bank that can be managed by client staff. Each of about a dozen live chatbots answers about $70\%$ of incoming questions without having to defer to a human for an answer. About two thirds of the automated guesses are confirmed by recruiters; the rest are corrected (Fig.~\ref{fig:dbcontrol}).

% We developed and validated the \texttt{nex-cv} metric on unbalanced multiclass classification with many categories: about 50-200 classes, with some classes having only 5-10 examples as well as classes with over a hundred. Each example used for training is a sentence or a phrase, either entered by a chatbot end-used, or provided by a chatbot project client as part of building an anticipated intent. Each dataset was mostly closed-domain: focusing on recruitment topics with some notable exceptions that are described further in the .

In ``Background'', we relate our work to  prior research on curated ML/NLP datasets and evaluation in chatbots. In ``Approach'', we describe the metric and provide its application and data context of use. In ``Validation Datasets'', we describe the datasets with which this metric has been validated.  In ``Validation'', we provide results from experiments conducted while developing and using the metric for over a year, addressing each of the needs of the metric, which make it a useful tool for multiple stakeholders in the chatbot design and maintenance process.

\begin{enumerate}
\itemsep0em 
    \item enable data quality improvements~(Fig.~\ref{fig:change})
    \item not be overly-optimistic~(Fig.~\ref{fig:humaneval})
    \item enable model-agnostic comparison~(Fig.~\ref{fig:compare})
\end{enumerate}

We contribute a metric definition, its validation with six real projects over the  course of one year (2018.Q2 through 2019.Q1), as well as an extensible implementation\footnote{\small{\texttt{http://github.com/jobpal/nex-cv}}}  and testing plan, which is described in  ``Metric Definition'' below.

\section{Background}

Chatbots, or ``text messaging-based conversational agents'', have received particular attention in 2010s \cite{jain2018evaluating}. Many modern text-based chatbots use relatively simple NLP tools \cite{abdul2015survey}, or avoid ML/NLP altogether \cite{pereira2018quality}, relying on conversation flow design and non-NLP inputs like buttons and quick-replies. Conversational natural-language interfaces for question-answering have an extensive history, which distinguishes open-domain and closed-domain systems ~\cite{mishra2016survey}. ML-based chatbots rely on curated data to provide examples for classes (commonly, ``intents''), and must balance being widely-accessible to many end-users, but typically specialized in the domain and application goal ~\cite{serban2015survey}. In practice, design  and development of a chatbot might  assume a domain more focused, or different, than real use reveals.

%  The first chatbot, ELIZA, was intended and as demonstration of superficiality of communication between humans and machines, but instead inspired enthusiastic reactions from people who ``unequivocally anthropomorphized it'' \cite{weizenbaum1976computer}.
%

In the chatbot  application  context, the training dataset  of a text classifier may be modified to improve that classifier's performance.  The classes\,---\,``intents''\,---\,are trained with synthetic data and constitute \textbf{anticipated}, rather than actual, use. Existing general-purpose platforms include this synthetic data step as part of design and maintenance~\cite{canonico2018comparison}. For example, when it comes to invocations for  a voice agent~\cite{ali2018crowdsourcing}, dataset construction encodes findings about how users might imagine asking for some action: the authors use a crowd-sourcing mechanism to achieve both consistency useful for classification, and reflection of user expectations in the dataset. We adopt a similar approach: enabling domain-experts (recruiters) to maintain the dataset helps map end-user (job-seeker) needs to recruiters' goals.

Data cleaning is not only relevant to chatbots. Model-agnostic  systems for understanding machine  learning can help iteratively  develop  machine learning models ~\cite{zhang2019manifold}. Developers tend to overlook data quality in favor of focusing on algorithmic improvements in building ML systems \cite{patel2008examining}. Feature engineering  can be made accessible to non-developers or domain experts, e.g. ~\cite{ribeiro2016should}.  We make use of representative examples in the process that surfaces \texttt{nex-cv} to non-developers; in describing this  process in  ``Metric Application'', we map it to the \textit{inspection-explanation-refinement}  process employed in~\cite{zhang2019manifold}. Enabling  non-developers to perform data cleaning effectively allows  developers  to  focus on model adjustments and feature engineering.

There are many ways to measure overall chatbot quality, such as manual check-lists of high-level feature presence ~\cite{kuligowska2015commercial, pereira2018quality}. Static analysis and formal verification may be used with a  specified flow ~\cite{porfirio2018authoring}. User behavior measurements\,---\,both explicit, like ratings or feedback, and implicit, like timing or sentiment\,---\,are  explored in ~\cite{hung2009towards}. During metric development,  we used qualitative feedback from domain-expert users, and key performance indicators (KPIs), such as automatic response rate. Regardless of overall evaluation approach, the use of a classifier as a component in a complex flow demands robust and actionable evaluation of that component.

\section{Approach}

The \texttt{nex-cv} algorithm selects some classes as plausible sources of negative examples, and then uses those to partition the given dataset into  training and test data (Alg.~\ref{algorithm:datapro}). Negative examples are useful in chatbot component evaluation: the end-user interaction with a chatbot is open-ended, so the system is expected to encounter input that it should recognize as outside its domain.

Low-membership classes are candidates for being ignored in training  and used as negative examples in testing. Two mutually-exclusive variations use the $K$ parameter for cutoff-based negative example selection (Alg.~\ref{algorithm:cutoff}); and the $P$ parameter  for proportional negative example selection (Alg.~\ref{algorithm:cutoff}). We focus on three settings, with $(K,P)$ set to $(0,0)$, $(0,0.15)$, and $(5,0)$. The values were tuned for typical distributions (see ``Validation Datasets''), and the  $(0,0)$ is a validating measure that is comparable to $5$-fold CV (see ``Metric Definition'').

We assume that low-population classes are all in the same domain as the rest. There may be exceptions: in some cases a new, small category may be created in response to new questions on an emergent topic well outside of the core domain. In our experience, this happens when there is a technical issue elsewhere on the site and the chatbot channel is used as an alternative to escalate the issue. In practice, our system handles this case well, even if the evaluation is less applicable. Such emergent categories either disappear over time: the topic is temporary; or grow: the topic becomes part of the domain.

\begin{algorithm}
\SetAlgoLined
\KwResult{$(X_{train},  y_{train}, X_{test},  y_{test})$}
 Require data $X, y$ s.t. $x_i$ is the input text that has gold standard label $y_i$ $\forall i$\;
 Require label sets $L_{SM}, L_{LG}$ s.t. $L_{SM} \cup L_{LG} = \{ y_i \mid y \}$
 Require test fraction $0 < t < 1$ and function $split_t(L)$ which randomly splits out two lists $L_1, L2$ s.t. $\frac{\vert L_2\vert}{\vert L \vert} =  t$ and $L_1 \cup L_2 = L$ \;

 \For{$L_j \in L_{LG}$}{
    $TR, TS = split_t({i | y_i \in y \land y_i==L})$\;
    $X_{train}, y_{train} \leftarrow x_i, y_i$ s.t. $i \in TR$ \;
    $TR, TS = split_t({i | y_i \in y \land y_i==L})$\;
    $X_{test}, y_{test} \leftarrow x_i, y_i$ s.t. $i \in TS$ \;
 }
 $TR_L, TS_L = split_t(\{j | y_j \in L_{SM}\})$\;
$X_{train}, y_{train} \leftarrow x_i, y_i$ s.t. $y_i \in TR_L$\;
  $X_{test}, y_{test} \leftarrow  x_i, \O$ s.t. $y_i \in TS_L$\;

 \caption{Negative Example Data Provision}
 ~\label{algorithm:datapro}
\end{algorithm}

\subsection{System Overview}

\begin{figure}
    \includegraphics[width=220px]{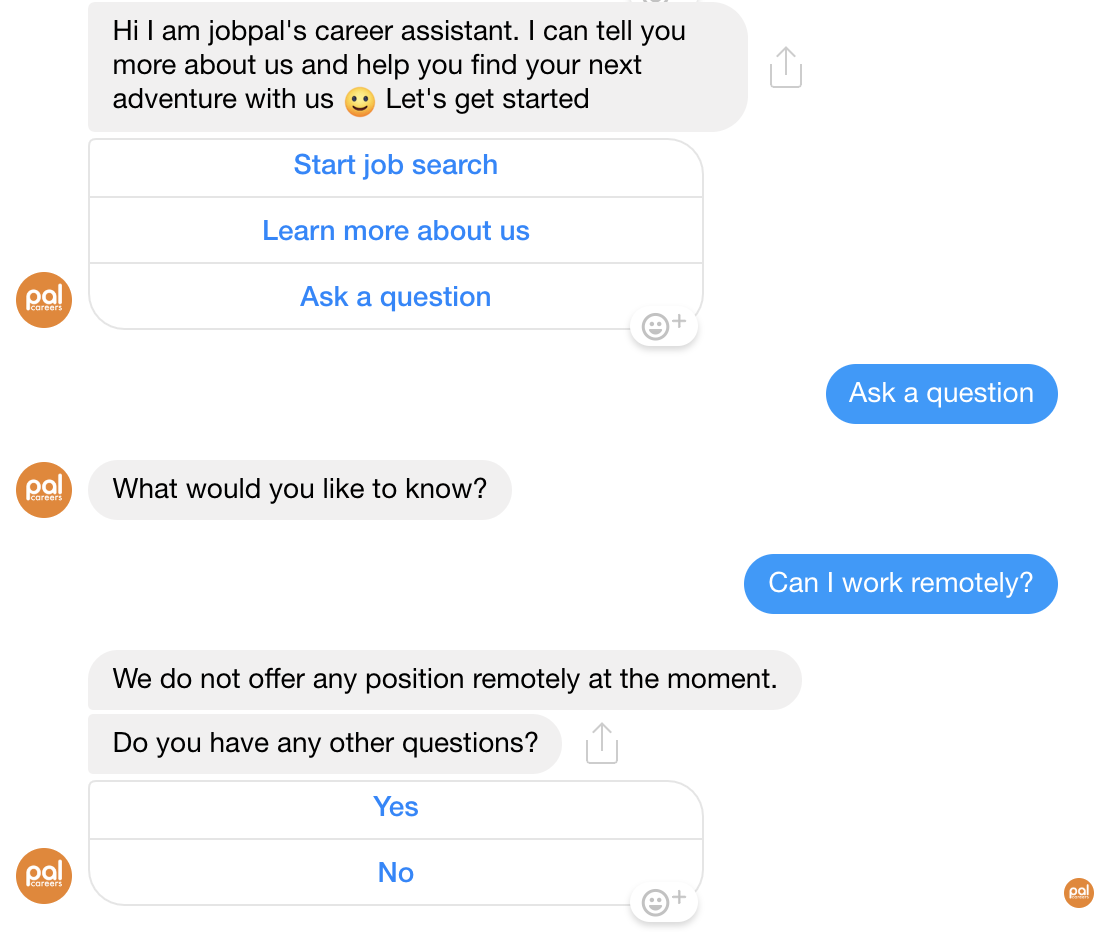}
    \caption{Here, the job-seeker's question receives an immediate answer, based on the ML/NLP classifier. If  confidence is too low, chatbot will defer to a human. \\ }
    \label{fig:sampledia}
\end{figure}

\begin{figure}
    \includegraphics[width=220px]{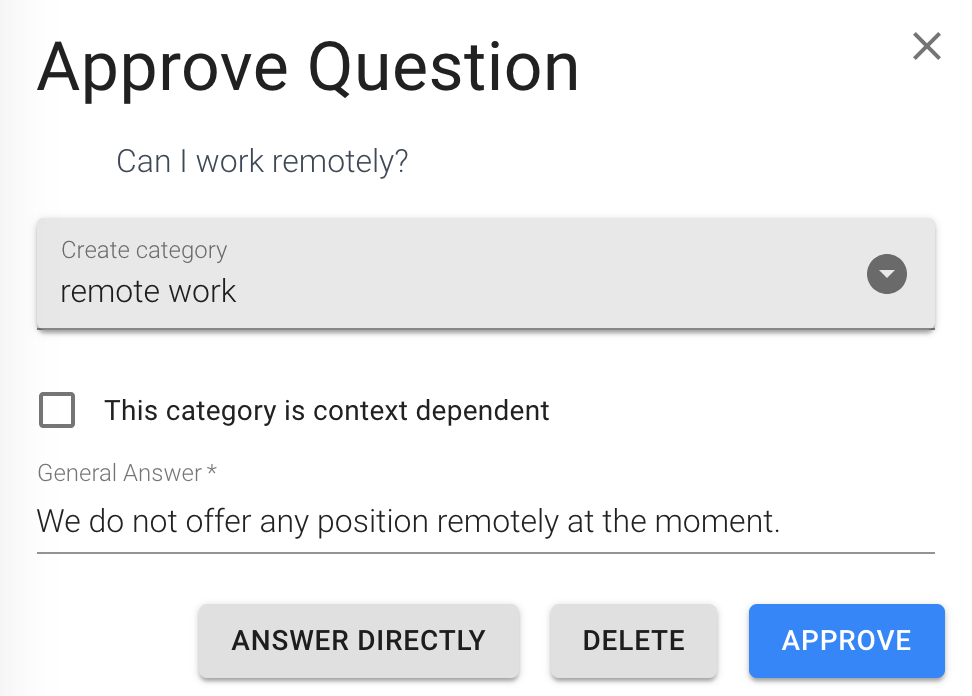}
    \caption{Even if the chatbot responds, recruiters can use a data management tool to review the answer.}
    \label{fig:dbcontrol}
\end{figure}

A chatbot (Fig.~\ref{fig:sampledia}) is based on two datasets (Fig.~\ref{fig:flowcontrol}), each maintained using a data management tool (Fig.~\ref{fig:dbcontrol}). Traffic varies widely between projects, but is typically  consistent  within a project. To provide a range: in one quarter in 2018, the highest traffic chatbot had about \textbf{$2000$} active users, of which about \textbf{$250$} (ca. $12\% $) asked questions. The lowest-traffic chatbot saw \textbf{$\tilde 65$} weekly active users, of which \textbf{$15$} (ca. $23\% $) asked questions. In both cases, a small number (2-4) of recruiters were responsible for maintaining the dataset.

The training set of the FAQ portion of each project contains between $1K$ and $12K$ training examples across between $100$ and $200$ distinct classes, usually starting with about $50-70$ classes and creating  new classes after the system goes live and new, unanticipated user needs are encountered. To build classifiers on datasets of this size, we use spaCy \cite{spacy2} and fastText  \cite{bojanowski2016enriching} for vectorization, with transformation for improved  performance \cite{arora2016simple}, and logistic regression with L2 regularization \cite{scikit-learn}. % Variations (such as tf-idf for feature extraction, or ensemble classification) are occasionally tested,  as are  language-specific improvements.

The  dataset for shared \textit{general intents} is maintained through the  data management tool by \textit{jobpal} staff. One such classifier is shared by all companies that use a particular language; projects span English, German, Chinese, and French. About $20$ general intents are trained with a total of about $1K$ to $1.5K$ training examples per language. These include intents that control the conversation (e.g., `stop', `help'). This shared  language-specific classification  step includes entity extraction of profession and city of interest to job-seekers; for  example, statements like `I want a [profession] job in [city]` and `do you  have any [profession] openings?' should all resolve to `job search' along  with  extracted keywords. Lastly, this classifier  also identifies very common questions that affect all chatbots\footnote{This was another outcome of the case study summarized in Fig.~\ref{fig:change}: we identified four categories of questions that we  could anticipate in all projects, but that  were  not in the expert domain of the FAQ, so we made modifications to the flow, the way the existing classifiers were  used, and the general intents training data, to help keep company-specific FAQ datasets more focused.}, but which are not in the recruitment domain: e.g.,  `how are you?' and `is this a robot?'.

The dialog  in Fig. ~\ref{fig:sampledia} shows the FAQ functionality of  the  chatbots, powered  by classification using  company-specific FAQ datasets (see also Fig.~\ref{fig:flowcontrol}). In most projects, users who ask question ask between 1 and 2 questions. The FAQ functionality is typically an addition to any existing information displays. Many of our chatbots also feature job discovery, including search and  subscriptions. Job search may be triggered by clicking on the button \texttt{[Look for a job]}, or writing something like ``I would like a \texttt{[profession]} job in \texttt{[location]}'' at almost any point in the  flow. If either of  location or profession is not specified, the user  is prompted, and the responses are used  to search current  openings, which are then shown. The user may submit  application or follow external links; the  user may also ask questions about  specific jobs or the employer more  generally. %We focus on the classifier component that powers the FAQ

\subsection{Metric Definition}

The code available online\footnote{\small{\texttt{http://github.com/jobpal/nex-cv}}} provides the evaluation implementation, an abstract black-box definition for a classifier, and two strategies to  help test an implementation. For integration testing, \texttt{CustomClassifier.test()} can be used to check consistency of classifier wrapper.  For functional testing,  \texttt{nex-cv} both $K=0$ (Alg. ~\ref{algorithm:cutoff}) and $P=0$ (Alg. ~\ref{algorithm:cutoff}) should yield comparable results  to $5$-fold cross-validation.

\begin{algorithm}
\SetAlgoLined
\KwResult{$L_{SM}, L_{LG}$}
 Require data $X, y$ s.t. $x_i$ is the input text that has gold standard label $y_i$ $\forall i$\;
 Require cutoff parameter $K > 0$ \;

 $L_{SM}=\{y_i \mid y_i$ in $y$, occurs $< K\}$ \;
 $L_{LG}=\{y_i \mid y_i$ in $y$, occurs $\geq K\}$ \;
 \caption{Cutoff Selection of Plausible Negative Example Classes}
  ~\label{algorithm:cutoff}

\end{algorithm}

In $k$-fold cross-validation, data is partitioned into $k$ sets of $(X_{train},  y_{train}, X_{test},  y_{test})$ such that $\frac{ \vert X_{test} \vert}{\vert X_{train}\vert } = 1/k$ (let the test fraction $t = 1/k$), and the training sets do not overlap. Then, each set of training data is evaluated using the corresponding test set. Evaluation can include many possible measures: accuracy or $F_1$; representative examples; confusion matrix; timing data; etc.

\begin{algorithm}
\SetAlgoLined
\KwResult{$L_{SM}, L_{LG}$}
 Require data $X, y$ s.t. $x_i$ is the input text that has gold standard label $y_i$ $\forall i$\;
 Require proportion parameter $0 \leq P < 1$ \;
 
   $L_{SM}=\{\}$ \;
   Let $Q= \{y_i \mid y_i \in y\}$, as queue sorted from least to most occurring in $X$ \;
     \While{$\frac{ \vert \{ i | x_i \in X \land y_i \in L_{SM} \} \vert }{\vert X \vert}$ $ < P$ }{
        Pop element $L$ from $Q$ \;
        $L_{SM}\leftarrow L$\;
     }
   $L_{LG}=\{ y_i \mid y_i$ in $ y$, not in $L_{SM} \}$ \;
 \caption{Proportional selection of Plausible Negative Example Classes}
  ~\label{algorithm:prop}

\end{algorithm}
In \texttt{nex-cv}, test  fraction $t$ is a setting ($0.2$  for all reported experiments), and data partitions may overlap. As shown in Alg.~\ref{algorithm:datapro}, representation of high-population classes is enforced. Then, low-population classes are also split using $t$, and included either in the training set with their ground truth label; or in the test set as a negative example. In practice, this results in about  $t$  of the data being in training. Some low-population classes in the training set should be included as this is representative of the dataset shape; many low-population classes  may affect the classification and confidence overall, depending  on classification  approach. Low-population classes are typically rare or relatively recent  topics, so interpreting them as plausible negative examples helps to test the classifier, and its measure of confidence.

\subsection{Validation Datasets}

\begin{figure}
    \includegraphics[width=220pt]{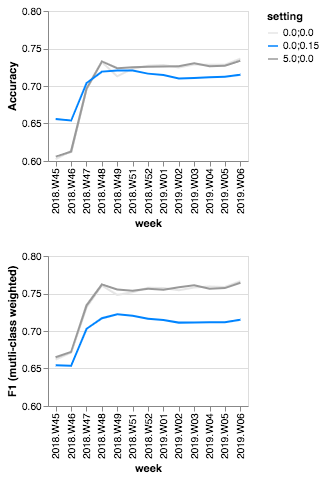}
    \caption{\textbf{Change in classifier performance as a result of data quality intervention.} Averages of daily 10-retry evaluations shown.} % The $0;0.15$ setting is not  overly-optimistic, relative to other measures.
    ~\label{fig:change}
\end{figure}

The seven datasets to which we report having applied the \texttt{nex-cv} metric are in the recruitment domain. Each dataset has about $50-200$ classes, and  most have classes with 5-10 members as well as classes with over a hundred. To characterize the content, we trained a classifier on an anonymous benchmark dataset \footnote{The clean, anonymized recruitment-domain-specific dataset in English was built by anonymizing and aggregating all FAQ datasets; using pairwise similarity between categories to group them. For an initial clustering, we used Jaccard index with a minimum of $0.09$, which balanced the goals of high coverage of example data ($~74$) and reasonable sizes of classes ($15$ examples per class); then,  this dataset was subject to iterative data quality  improvements as described  further and exemplified in Fig.~\ref{fig:change} until  a final set of about $800$ examples over about $47$ categories was developed. This initial domain-specific clustering was performed on English, but has since been extended  to other supported languages; the results reported are specific to English,  however.} and used it to classify a random recent sample of $6K$  English-language  questions.

About $25\%$ of recent end-user queries in English fall into $5$ categories: (1) Application Process; (2) Salary; (3) Professional Growth and Development; (4) Internships; (5) Contact a Human.

Another $25\%$ of  end-user queries fall into $14$ categories: Application Evaluation; Application Deadline; Application Delete or Modify; How Long to Apply and Hear Back; Qualification; Application Documents; Language Expectations; Thesis; Working Hours; Location; Starting at the Company; Commute; Equipment; Benefits.

About $40\%$ of overall requests were not recognized (with a confidence of $0.5$ or  higher) as any of the categories in the anonymous benchmarking set. Upon manual inspection, some of these test questions were noise,  and many were topics specific  to particular  company FAQs, such as concerning specific work-study programs; details of the application software; and other more niche topics.

The classification datasets share some overlapping topics; each also has a specific set of additional topics. Each dataset has the typical shape of a few larger classes, and many smaller ones, which have an indirect relationship to what data is expected. The use of low-population classes as plausible negative examples takes advantage of both the content of the data (closed-domain, with a topic-specific core but a considerable number of additional, outlying topics) and the membership distribution of the classes (a few well-populated  ones, and  many much smaller classes).

The \texttt{nex-cv} metric may apply  in other problems or domains, but we developed and validated it in a series of experiments with six live datasets, in English and German (see Fig.~\ref{fig:humaneval}, of which chatbot E is also the subject of Fig.~\ref{fig:change}), in addition to the seventh aggregate anonymous benchmark dataset described above,  which was used for the comparison in Fig.~\ref{fig:compare}.

%within the context of chatbot application. The application context also The expectation of the domain-expert  users of the system who encode anticipated chatbot use into these curated datasets also implies that the classifier is not quite used directly:  any text match must always fetch the class it  is associated with, even if the trained classifier may be giving a different answer. This context of use also makes the  \texttt{nex-cv} metric more applicable, because 

\section{Validation}

\begin{figure*}
    \includegraphics[width=450pt]{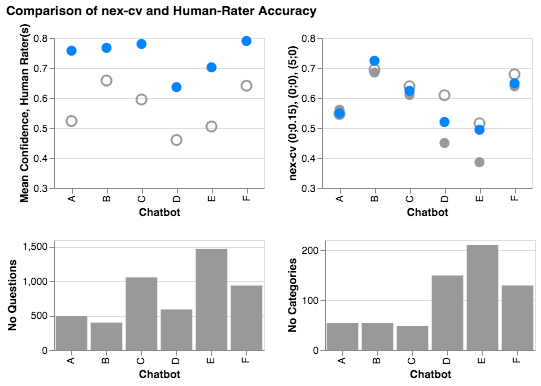}
    \caption{\textbf{Comparison of \texttt{nex-cv} and Human-Rater Accuracy}. The six datasets from pseudonymous chatbots tested had a different number of questions (examples) and categories (classes), as shown in the bottom row. The human-rater estimate of accuracy (top left, blue) is consistently more lenient than any of the automated measures (top right). The $(0;0.15)$ setting (top right, blue) is not consistently more or less optimistic than the other settings.}
    ~\label{fig:humaneval}
\end{figure*}

\begin{figure*}
    \includegraphics[width=450pt]{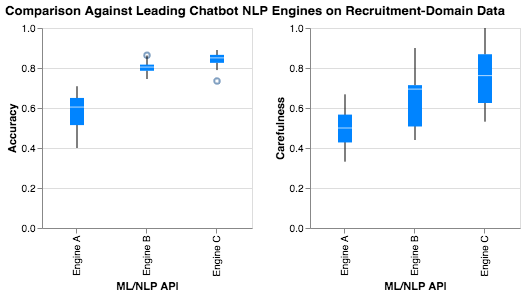}
    \caption{\textbf{Comparison Against Leading Chatbot NLP Engines on Recruitment-Domain Data.} Engine C wraps jobpal's system; Engines A and B wrap external general-purpose chatbot platforms.}
    ~\label{fig:compare}
\end{figure*}

The following case studies  validate the metric relative to each of  three requirements: (1) enable data quality improvements, as in Fig.~\ref{fig:change}, (2) not be overly-optimistic,  as in Fig.~\ref{fig:humaneval}, (3) enable model-agnostic comparison, as in Fig.~\ref{fig:compare}.

%The role of curation and maintenance of the dataset used for training in the organizational process has two implications for what it means for a classifier performance metric to be usable. First, because the membership in classes reflects anticipation of use or curation of ``clean'' data, the training dataset class membership may not be representative of what is actually encountered. Second, performing $k$-fold cross-validation also does not account for behavior with unseen data, particularly if the fallback behavior if fallback human escalation is not implemented as a separate class. We present description and validation of the metric

\subsection{Metric Application}

The goal of usefulness includes interpretability: ``provid[ing] qualitative understanding between the input variables and the response... [taking] into account the user’s limitations in~\cite{ribeiro2016should}. Usefulness combines this with actionable support  of chatbot design. The users include, in this case, non-developer staff on both vendor and client side: recruiters and project managers.

Through iteration on internal tools, we found that displaying performance information in the form of,  ``which 2-3 topics  are the biggest problem?'' was most effective for understanding, communication, and action. Over the course of a year, the \texttt{nex-cv} metric informed this analysis. During this time, both qualitative feedback and KPIs have validated that it was effective both for trust and for the end-user experience. The \textit{automation rate} KPI\,---\,proportion of incoming queries that did not need deferral to a human, but answered immediately,  as in Fig.~\ref{fig:sampledia}\,---\,has risen to and remained at $70-75\%$ across projects mainly\footnote{The data training UI design contributes to data quality; in the months following the intervention shown in Fig.~\ref{fig:change}, the UI was redesigned to  address outstanding usability  problems, with very positive feedback from domain-expert  users. A more in-depth discussion of  the role of human factors in human-in-the-loop systems  is out of scope for this paper.} due to data quality support during both design and maintenance.

In one illustrative project (Fig.~\ref{fig:change}) the  automation  rate had  become as low as $40\%$. The recruiters responsible for dealing with escalated questions became frustrated to see questions come up that had been asked before. Action needed to be  taken, and this project became one of the first case studies for developing the application of  \texttt{nex-cv} internally. After intervention, automated response rate rose into the  desirable $70$s range and remained. The quality improvements   were explained and implemented by an internal project manager,  who pro-actively included client domain-expert users in explanations over calls and emails over what improvements were made and why.  Initially, $~200$ classes were trained with $~1K$ examples, with long tail of low-population classes. Following intervention, dataset grew by $25\%$ and, despite concept drift risk, did not deteriorate.

To use \texttt{nex-cv}, we aggregate the confusion matrix from the $K=0;P=0.15$ setting  and rank how confused a \textit{pair} of categories is. The most confused 2-3 pairs of classes are then the focus of conceptual, manual review in the dataset. Evaluation is performed again, producing a new ranking that guides the next 2-3 classes to focus on, until the metric falls below an acceptable threshold. There are other sources of classification error, but overlap between conceptually related pairs of classes accounts for most of the data quality problems we encounter in the datasets in practice, and are particularly understandable than other forms of error. This relatively simple approach is implemented as a Jupyter notebook accessible to non-developers (internal project  managers).

The details of pairwise measures and acceptability threshold were developed iteratively based on project  manager feedback. The project managers also honed processes and intuitions for communicating this information to clients effectively. In extreme situations  as that shown in Fig.~\ref{fig:change} the project  managers made a presentation to get buy-in and implemented data quality improvements on their own. However, the typical practice now is  to provide explanations, in calls and emails, of the ``confusions'' between one or few pairs of specific categories to the client. This practice builds awareness of data quality across stakeholders,  and the domain-experts (recruiters) are better able to use the system to create the envisioned chatbot functionality without major intervention. As the number of projects grows,  the metric can  be used by project managers to monitor and prioritize data quality  improvement tasks.

\subsection{Metric is not Overly Optimistic}

% Typically accuracy is not used, in favor of F1, for classification. In the binary case, F1 is defined as TK; extension to multi-class classification problems may mean either weighted or unweighted by class population. Weighted F1 is not appropriate in this case: it assumes that the unequal distribution TK

One of the practical motivations for a new metric was the sense that the existing metrics were too optimistic to be useful to improve chatbot behavior in response to overall qualitative feedback. As shown in Fig.~\ref{fig:change}, for example, the typical $F_1$ metric is more optimistic than \texttt{nex-cv}.

As an initial step of validating the metric, we applied it in the case of six under-performing datasets that required some intervention. Fig.~\ref{fig:change} shows the differences in data abundance and classifier quality across these six pseudonymized snapshots. Internal QA staff gave the human rating scores by considering whether a question-answer pairs seemed reasonable: they could pick ``yes'' ``no'' and ``can't tell''; in most cases, the appropriateness was not ambiguous. As shown in Fig.~\ref{fig:humaneval}, the human-rater estimate of quality is consistently more lenient than any of the automated measures. The Chatbot E in this case is the same project as shown in Fig.~\ref{fig:change}, prior to improvements.

Four of the  six datasets analyzed had a very big difference  between the human estimate of  quality and the automated estimate, which, upon investigation, revealed that there were significant conceptual overlaps in the classes that the recruiters had trained, and the answers given. So, indeed, the classifier was making surprisingly adequate guesses,  but which were very low-confidence. Following the intervention described in the previous section,  which includes ongoing communication of any outstanding  problems by project  managers to  recruiter teams, this type of error became rare and quickly-addressed.

\subsection{Metric can be used for Internal and External Comparison}

We used the \texttt{nex-cv} metric to help compare the performance of our classification  component with  two leading vendors for general-purpose chatbot development. Fig.~\ref{fig:compare} shows the comparison between \textit{jobpal} and 2 leading vendors in the space. The three  settings of the  metric\footnote{Where $(K,P)$ are $(0,0)$, $(0,0.15)$, and $(5,0)$, respectively, as differentiated in both Fig.~\ref{fig:change} and Fig.~\ref{fig:humaneval}.} were aggregated to provide a plausible range of estimated performance. The range of accuracy was significantly higher  for our domain-specific classifier, than those trained using general-purpose tools.

Aside from being useful to classify into known classes,  the metric must account  for fallback or escalation. This may be modeled as a separate class (as one of the external engines does with the ``fallback'' intent), or by relying on confidence scores from classifiers that produce measures of confidence (all engines provide some estimate of confidence that may be used). The ``carefulness'' score was included to represent how useful  the confidence score is for  deciding when to decline an answer: the number of incorrect  guesses that  were rejected due to too-low confidence scores divided  by total no-answer-given cases (no guess or low-confidence guess).

Fig.~\ref{fig:compare} shows that the performance of our  ML/NLP  component on our  domain-specific dataset is better than that of two popular general-purpose platforms, both in terms of classification accuracy, and rate of deferral due to low-confidence answers. This comparison mechanism validates our system relative to existing external services in a way that is interpretable by various internal stakeholders, not only the developer staff.

\section{Conclusion}

We described and validated the \texttt{nex-cv} metric, which is a modification of cross-validation that makes use of plausible negative examples from low-population classes in the datasets typical of our application area and domain.

Existing chatbot guidelines leave error handling to the designer: ``transparency'' is included as an important topic \cite{dialogflowerrors}, but, in practice, why something does not work, and under what conditions, can puzzle designers and developers, not just end-users. We presented  on a metric that can be used  by a variety of relevant stakeholders to  understand, communicate, and improve text classifier performance by improving data quality.

In  future work, we aim to  explore other text classifier  and chatbot evaluation strategies, keeping in mind the needs for understandability and transparency in this multi-stakeholder design process and  maintenance  practice. 
% TK explore the applicability of this metric in other text classification  domains, as ell as experiments to more  rigorouslt explore the data quality improvement process, which currently produces good end-outcomes and qualitative  anecdotal feedback from all stakeholders, but by more rigorously investigating the information visualization and ML summarizatin things we could se. ALSO this is basically  topic with short next,  but would be interesting to apply also in sentiment analysis. 

%\section*{Acknowledgments}

%The acknowledgments should go immediately before the references.  Do
%not number the acknowledgments section. Do not include this section
%when submitting your paper for review. \\

\bibliography{naaclhlt2019}
\bibliographystyle{acl_natbib}

\end{document}